\documentstyle[multicol,aps]{revtex}

\begin{document}
\title{Observable Dirac electron in accelerated frames}
\author{Marc-Thierry Jaekel$^a$ and Serge Reynaud$^b$}
\date{March 1999}
\address{$(a)$ Laboratoire de Physique Th\'eorique de l'ENS 
\thanks{Laboratoire du CNRS associ\'e \`a l'Ecole Normale 
Su\-p\'e\-rieu\-re et \`a l'Uni\-ver\-si\-t\'e Paris-Sud},\\
24 rue Lhomond, F75231 Paris Cedex 05 France \\
$(b)$ Laboratoire Kastler Brossel 
\thanks{Laboratoire de l'Ecole Normale Su\-p\'e\-rieu\-re et de
l'Uni\-ver\-si\-t\'e Pierre et Marie Curie asso\-ci\'e au CNRS}, \\
UPMC case 74, 4 place Jussieu, F75252 Paris Cedex 05 France}
\maketitle

\begin{abstract}
We present a new quantum algebraic description of an electron localized in
space-time. Positions in space and time, mass and Clifford generators are
defined as quantum operators. Commutation relations and 
relativistic shifts under frame transformations are determined
within a unique algebraic framework. Redshifts, {\it i.e.} 
shifts under transformations to uniformly accelerated frames, are
evaluated and found to differ from the expressions of classical
relativity.

\end{abstract}

\begin{multicols}2

\section{Introduction}

Well-known difficulties affect Quantum Field Theory at its interface
with mechanical, inertial or gravitational effects.
In particular, the problem of quantum systems in accelerated frames
is quite poorly mastered when compared with quantum theory on one hand
and classical relativity on the other hand.
Prospects for improved tests of the effect of gravitation on atomic
clocks however make this question inescapable. 
We shall show below that the standard theoretical framework, despite
its well-known deficiencies with respect to this problem, may be
reformulated to allow for the definition of observables describing 
localization of a quantum system in space-time. Here we shall
consider the exemplary quantum system constituted by a single
electron.
 
Special relativity has led to a profound revision of our conception of
space-time. Basic elements of the new conception are events localized in
space and time. Time definition and clock synchronization
correspond to dating of events such as, respectively, ticks of a clock and
emission or detection of light pulses \cite{Einstein05}. Positions in
space-time of such events are physical observables which differ
from coordinate parameters on a space-time map. Consequently, relativistic
shifts of observables under frame transformations are related to
relativistic symmetries and are distinct from mere changes of 
coordinate mapping.

These points which are already delicate in classical relativity still
raise more acute difficulties in a quantum context. 
To illustrate these difficulties, let us consider the simple case of 
an electron. Clearly, its position in space is described by $3$ 
quantum observables conjugated to momentum components. 
Lorentz transformations of these positions cannot be properly 
understood unless a position in time is also defined as a quantum
observable \cite{Schrodinger30}. But such a time operator is commonly
considered to be unavailable in the standard quantum formalism
where time remains a classical parameter used to describe evolution.
More generally, the definition of space-time observables fulfilling
quantum and relativistic requirements has to face the problem
that the changes of coordinate mapping of relativity have the status of
gauge transformations in quantum theory and, as a result, 
that they cannot affect physical observables \cite{deWitt63,Rovelli91}.

The answer to these problems, imported from relativistic theories,
is to associate the definition of observables as well as the evaluation 
of their shifts with the algebra of symmetries.
In this respect, Einstein's synchronization or localization procedures 
are exemplary questions by their direct relations with the symmetry 
properties of electromagnetic fields. Relativistic effects on space-time
positions are well-known to reflect invariance of the laws of physics 
under Lorentz frame transformations. Moreover, Einstein's localization
relies on Maxwell equations which are invariant also 
under dilatations and conformal transformations to uniformly accelerated 
frames \cite{Bateman09,Cunningham09,FultonRW62}. 

These symmetry principles still hold in a quantum context \cite{BinegarFH83}. 
It is then possible to define observables associated with position in
space-time of events and to evaluate their relativistic shifts.
For electromagnetic fields, an event may be defined as the
intersection of two pulses and its position in space-time may be written 
from the generators of Poincar\'{e} transformations and of dilatations. 
These position observables are found to be conjugated to momentum 
operators while simultaneously obeying explicitly Lorentz invariance.
A time observable is found besides space ones and 
the shifts of these observables under Lorentz transformations 
conform to expectations of classical relativity \cite{JaekelR96}.

Furthermore, redshifts, {\it i.e.} relativistic shifts under transformations 
to uniformly accelerated frames may be evaluated in the same manner
from conformal symmetry. Localization observables are defined for field
states which correspond to a non vanishing mass since they contain 
photons propagating in different directions. Remarkably, this mass observable 
experiences a redshift which reproduces exactly the effect of the gravitational 
potential arising in accelerated frames according to Einstein equivalence 
principle \cite{JaekelR97}. Hence, conformal symmetry forces the mass 
unit to scale as the inverse of the space-time unit and therefore 
corresponds to preservation of the quantum constant $\hbar $ under 
frame transformations \cite{Dicke62,Sakharov74,Hoyle75}.

These results have been derived for electromagnetic quantum systems.
They may certainly be expected to have a more universal character.
As an important example, an electron-positron pair is a system which may
decay into a pair of photons. The position of the decay 
reduces to the point of coincidence of the two emitted photons.
Meanwhile, the mass is conserved in the annihilation process.
It should therefore be expected that the redshift law consistent with
Einstein equivalence principle is valid not only for the post-decay
electromagnetic state but also for the pre-decay electron-positron pair. 
But to clear up this question, one must also be able to define
localization observables for massive quantum systems such as electrons.
Dirac theory of electrons \cite{Dirac28} is certainly insufficient for 
that purpose since the coordinate parameters used to write quantum 
fields associated with electrons cannot be considered as localization
observables.

For massless field theories such as electromagnetism, positions have 
been built on Poincar\'e and dilatation symmetry and redshift
laws on the full conformal symmetry. The same achievements cannot be 
extended for electron as long as the latter is described by field
theories, like Dirac theory, which violate dilatation or conformal 
symmetry by treating electron mass as a classical constant.
But modern descriptions of electron consider that its 
mass is generated through an interaction with Higgs fields 
\cite{ItzyksonZch12}. 
A quantum representation of electron mass is in fact inescapable since 
it is, at least partly, generated by electromagnetic self-energy.
It must therefore present intrinsic quantum fluctuations \cite{JaekelR93}. 
Being a quantum operator rather than a classical parameter,
it should vary under frame transformations. 
Standard forms of coupling to the Higgs field obey conformal invariance 
\cite{Pawlowski98}, so that the mass redshift may be expected to follow 
the same law as for electromagnetic fields and, therefore, to fit
Einstein equivalence principle. 

The aim of the present paper is to build up a new quantum description 
of electrons obeying conformal symmetry and fulfilling the expectation 
of the preceding paragraph. We consider that electrons are described by a
conformally invariant field theory which we do not specify in more detail.
Using general properties of conformal algebra and a few
specific assumptions, we write down localization observables for 
electrons and deduce their quantum and relativistic properties. 
The specific assumptions are drawn from the phenomenology of
electrons which are fermions with a spin $\frac1 2 $. They are
written in a purely quantum algebraic manner and thus present 
analogies with Connes' non commutative geometry \cite{Connes94}.
These analogies will be discussed as well as 
differences between the two approaches.

\section{Hermitian localization observables}

Localization observables will be built upon the algebra of symmetries in the
manner already used for electromagnetic fields \cite{QED98,EPJD98}. 

Firstly this algebra
contains Poincar\'{e} algebra characterized by the following relations 
\begin{eqnarray}
&&\left( P_\mu ,P_\nu \right) =0\qquad \left( J_{\mu \nu },P_{\rho
}\right) =\eta _{\nu \rho }P_\mu -\eta _{\mu \rho }P_\nu   \nonumber \\
&&\left( J_{\mu \nu },J_{\rho \sigma }\right) =\eta _{\nu \rho }J_{\mu
\sigma }+\eta _{\mu \sigma }J_{\nu \rho }-\eta _{\mu \rho }J_{\nu \sigma
}-\eta _{\nu \sigma }J_{\mu \rho }  \label{PJ}
\end{eqnarray}
$P_\mu $ and $J_{\mu \nu }$ are the components of energy-momentum vector
and angular momentum tensor associated with a single electron. Commutators 
(\ref{PJ}) suffice to describe quantum and relativistic properties.
They determine characteristic commutation relations of the generators
considered as quantum observables, according to the following notation 
\begin{eqnarray}
\left( A,B \right) &\equiv& \frac {AB-BA} {i\hbar}
\nonumber
\end{eqnarray}
Meanwhile they represent the relativistic shifts of observables under
translations and rotations. The Minkowski tensor 
\begin{eqnarray}
\eta _{\mu \nu } &=& {\rm diag} \left( 1,-1,-1,-1 \right) 
\nonumber
\end{eqnarray}
is used throughout the paper for manipulating indices.

We then assume that the symmetry algebra contains a generator $D$ which
generates relativistic shifts under global dilatations according to the
conformal weight of observables 
\begin{equation}
\left( D,P_\mu \right) = P_\mu \qquad \left( D,J_{\mu \nu }\right) = 0
\label{PJD}
\end{equation}
Equation (\ref{PJD}) constitutes a key assumption of our approach to the
problem of localization observables. It is the source of important
differences with Dirac theory of electrons although, as we shall see later
on, it leads to close analogies with the latter. 
To illustrate these differences, we consider the mass observable 
$M$ defined in accordance with the standard relativistic relation 
\begin{equation}
M^2 = P^2 \equiv P^\mu P_\mu   
\label{defM2}
\end{equation}
as a Lorentz scalar with the same conformal weight as momenta
\begin{equation}
\left( P_\mu , M \right) = \left( J_{\mu \nu}, M \right) = 0
\qquad \left( D, M \right) = M 
\label{PM}
\end{equation}
These relations determine the mass $M$ up to an ambiguity which will
be cleared up later on. At the moment, it is clear that any classical 
treatment of $M$ would lead to a vanishing commutator with $D$
and, therefore, to a contradiction with (\ref{defM2}). On the contrary, 
the commutator $\left( D,M\right)$ written in (\ref{PM}) is necessary 
in any framework where mass has its proper conformal dimension. 

Localization observables representing positions in space-time 
may be built on Poincar\'{e} and dilatation generators. 
First, spin observables are introduced in a relativistic framework
through the Pauli-Lubanski vector \cite{ItzyksonZch2} 
\begin{eqnarray}
W_\mu  &\equiv &-\frac1 2 \epsilon _{\mu \nu \rho \sigma }J^{\nu \rho
}P^{\sigma }  \nonumber \\
\left( P_\mu ,W_{\rho }\right)  &=&0\qquad \left( J_{\mu \nu },W_{\rho
}\right) =\eta _{\nu \rho }W_\mu -\eta _{\mu \rho }W_\nu 
\end{eqnarray}
$\epsilon _{\mu \nu \lambda \rho }$ is the completely antisymmetric Lorentz
tensor 
\begin{eqnarray}
\epsilon _{{\rm 0123}} &=&-\epsilon ^{{\rm 0123}}=+1  \nonumber \\
\epsilon _{\mu \nu \rho \sigma } &=&-\epsilon _{\mu \nu \sigma \rho
}=-\epsilon _{\mu \rho \nu \sigma }=-\epsilon _{\nu \mu \rho \sigma }
\nonumber
\end{eqnarray}
Commutators between components of the spin vector define a spin tensor 
\begin{equation}
S_{\mu \nu } = \frac {\left( W_\mu ,W_\nu \right)} {M^2} 
= \epsilon _{\mu \nu \rho \sigma }  \frac {W^\rho P^\sigma } {M^2} 
\label{defS}
\end{equation}
Spin observables are transverse with respect to momentum 
\begin{equation}
P^\mu S_{\mu \nu }=P_\mu W^\mu =0
\end{equation}
The square modulus of the Lorentz vector $W^\mu $ is a Lorentz scalar that
we can write under its standard form in terms of a spin number $s$ taking
integer or half-integer values 
\begin{equation}
\frac{W^2 }{M^2 }=-\hbar ^2 s\left(s+1\right) 
\end{equation}
The spin number will be fixed to the value $\frac 1 2$ in the following.
Throughout the paper, the velocity of light is set to unity while 
the Planck constant $\hbar $ is kept as the characteristic 
scale of quantum effects.

We are then able to define position observables as the quantities $X_\mu $ 
solving the following equations 
\begin{equation}
J_{\mu \nu }=P_\mu \cdot X_\nu -P_\nu \cdot X_\mu +S_{\mu \nu
}\qquad D=P^\mu \cdot X_\mu 
\label{eqJDX}
\end{equation}
The dot symbol denotes a symmetrised product for non commuting observables 
\begin{eqnarray}
A\cdot B &\equiv& \frac{AB+BA}2 
\nonumber
\end{eqnarray}
Position observables are then obtained as 
\begin{equation}
X_\mu =\frac{P_\mu }{M^2 }\cdot D+\frac{P^{\rho }}{M^2 }\cdot J_{\rho
\mu }  \label{defX}
\end{equation}
They are shifted under translations, dilatation and rotations exactly as
ordinary coordinate parameters are shifted under the corresponding
transformations in classical relativity 
\begin{eqnarray}
&&\left( P_\mu ,X_\nu \right) = -\eta _{\mu \nu } \qquad \left( D,X_{\mu
}\right) =-X_\mu   \nonumber \\
&&\left( J_{\mu \nu },X_{\rho }\right) = \eta _{\nu \rho }X_\mu -\eta _{\mu
\rho }X_\nu   \label{PX}
\end{eqnarray}
In particular, the first of these relations means that positions 
are conjugated to momenta. However, various position components do not commute 
\begin{equation}
\left( X_\mu ,X_\nu \right) =\frac{S_{\mu \nu }}{M^2 }  \label{XX}
\end{equation}
Note that relations (\ref{defS},\ref{defX},\ref{XX}) are well defined only
when the squared mass $M^2$ differs from $0$. For electromagnetic
states, such a condition was revealing that a localized event may be defined
only with photons propagating in at least two different directions \cite
{QED98,EPJD98}. This problem does not arise for an electron which has  
a non vanishing mass. 

Relation (\ref{XX}) clearly indicates that the
conceptions of space-time inherited from classical relativity have to be
revised for quantum objects. Electron cannot be
considered as a sizeless point but rather as a fuzzy spot with a size 
of the order of Compton wavelength.

\section{Canonical localization variables}

In this context, it is a remarkable and useful property that an auxiliary 
set of variables may be defined
which possesses algebraic properties of canonical variables.

To obtain these observables, we first consider the involutive duality
correspondance 
\begin{equation}
\tilde{S}_{\mu \nu }=\frac{i}2 \epsilon _{\mu \nu \rho \sigma }
S^{\rho \sigma} 
\qquad S^{\mu \nu }=\frac{i}2 \epsilon ^{\mu \nu \rho \sigma } \tilde{S}_{\rho\sigma }
\end{equation}
This definition is such that 
\begin{equation}
\tilde{S}_{\mu \nu }=i\frac{P_\mu W_\nu -P_\nu W_\mu }{M^2 }
\qquad P^\mu  \tilde{S}_{\mu \nu }=i W_\nu 
\end{equation}
We then introduce a self-dual representation of spin 
\begin{eqnarray}
s_{\mu \nu }&=& S_{\mu \nu }+\gamma \tilde{S}_{\mu \nu }
\nonumber \\ 
\tilde{s}_{\mu \nu }&=& \frac{i}2 \epsilon _{\mu \nu \rho \sigma }
s^{\rho \sigma} = \gamma s_{\mu \nu }
\label{defs}
\end{eqnarray}
where $\gamma $ is a dimensionless Lorentz scalar with a square equal to
unity 
\begin{equation}
\left( P_\mu ,\gamma \right) =\left( J_{\mu \nu },\gamma \right) =\left(
D,\gamma \right) =0\qquad \gamma ^2 =1
\end{equation}
As a consequence of self-duality (\ref{defs}), $\gamma $ is 
the orientation of the canonical spin tensor $s_{\mu \nu }$
with the two eigenvalues $\pm 1$ associated with the right/left components.
It plays the same role here as $\gamma_5 $ in Dirac theory \cite{ItzyksonZapA2}. 

A new definition of position variables is associated with
the self-dual spin tensor (\ref{defs})
\begin{eqnarray}
&&x_\mu  = X_\mu  - i\gamma \frac{W_\mu}{M^2}
= X_\mu  - \frac{P^\nu s_{\nu \mu }}{M^2}  
\nonumber \\
&&J_{\mu \nu } = P_\mu \cdot x_\nu - P_\nu \cdot x_\mu 
+ s_{\mu \nu} \qquad D=P^\mu \cdot x_\mu
\label{defx}
\end{eqnarray}
Poincar\'{e} and dilatation generators take the same form 
in terms of both sets of variables (\ref{eqJDX}) and (\ref{defx}).
The variables $x_\mu $ and $s_{\mu \nu }$ are quantum operators as 
$X_\mu $ and $S_{\mu \nu }$ but they obey
canonical commutation relations 
\begin{eqnarray}
&&\left( P_\mu ,x_\nu \right) =-\eta _{\mu \nu }\qquad \left( x_\mu
,x_\nu \right) =0  \nonumber \\
&&\left( s_{\mu \nu },s_{\rho \sigma }\right) =\eta _{\nu \rho }s_{\mu
\sigma }+\eta _{\mu \sigma }s_{\nu \rho }-\eta _{\mu \rho }s_{\nu \sigma
}-\eta _{\nu \sigma }s_{\mu \rho }  \nonumber \\
&&\left( P_\mu ,s_{\nu \rho }\right) =\left( x_\mu ,s_{\nu \rho }\right)
=0  \label{Px}
\end{eqnarray}
The localization observables $X_\mu $ and $S_{\mu \nu }$ obey hermiticity
conditions, even if they are not self-adjoint \cite{QED98,Toller98}.
Relations (\ref{defs},\ref{defx}) thus show that canonical variables $x_{\mu
}$ and $s_{\mu \nu }$ are not hermitian. This is an important output of
our quantum approach to the localization problem. One may define either 
hermitian observables with non canonical commutation relations or canonical 
variables which lead to simpler explicit calculations but are not hermitian. 

{}From now on, we focus our attention on canonical variables. They can be seen as
quantum algebraic versions of the position parameters and spin matrices of
Dirac theory, as it will become clear in forthcoming computations.
We emphasize that $4$ positions in space and in time have been defined 
in contrast with previous studies of the localization problem where only 
positions in space were introduced \cite{Pryce48,Fleming65}. This
means that the requirement enounced by Schr\"{o}dinger \cite{Schrodinger30}
has been met: Lorentz transformations may now be properly described within 
quantum theory. Precisely, canonical positions are quantum observables which
are transformed according to classical laws of special relativity
\begin{eqnarray}
&&\left( P_\mu ,x_\nu \right) =-\eta _{\mu \nu }\qquad \left( D,x_{\mu
}\right) =-x_\mu   \nonumber \\
&&\left( J_{\mu \nu },x_{\rho }\right) =\eta _{\nu \rho }x_\mu -\eta _{\mu
\rho }x_\nu 
\end{eqnarray}

\section{Clifford algebra}

We narrow still more the scope by considering electrons which are
fermions with a spin number $s=\frac 1 2 $.

We notice that the involution $\gamma$ commutes with the squared mass 
$M^2=P^2$ while $M$ commutes with $\gamma^2=1$. 
These conditions are fulfilled as soon as $\gamma$ and $M$ commute or
anticommute. We will assume in the following that $\gamma$ and $M$ are 
anticommuting variables.
This entails that $M$ may be written as follows
\begin{eqnarray}
&&M = \varepsilon \left| M\right| \qquad \left| M \right| = \sqrt{P^2}
\nonumber \\
&&\left( P_\mu ,\varepsilon \right) =\left( J_{\mu \nu }, \varepsilon \right) 
=\left( D, \varepsilon \right) = 0 
\nonumber \\
&&\varepsilon ^2 =1 \qquad \gamma \cdot \varepsilon = 0 
\label{anticom}
\end{eqnarray}
The modulus $\left| M \right|$ is equal to the norm of the energy-momentum
vector while the sign $\varepsilon$ is a further dimensionless Lorentz 
scalar with a square equal to unity. 
Clearly, the two eigenvalues $\pm 1$ of $\varepsilon $ are associated
with the electron/positron components so that $\varepsilon $ 
corresponds to charge in Dirac theory \cite{ItzyksonZch2}.
The fact that $\gamma$ and $\varepsilon$ anticommute or
that $\gamma$ and $M$ anticommute is an important property. It means that 
the orientation $\gamma $ is an operator changing the mass sign 
$\varepsilon$ into its opposite or, equivalently, that $\varepsilon$ 
interchanges the two spin orientations. We show in the following that this
property is sufficient, when taken together with the general properties of 
conformal algebra, to build up a quantum algebraic theory of electrons.

Velocity observables $V_\mu$ may be defined as ratios of momenta 
$P_\mu$ to mass $M$ or, equivalently, by applying the derivation operator 
$\left( \ ,M\right) $ to hermitian positions $X_\mu$  \cite{FoP98} 
\begin{equation}
V_\mu = \frac{P_\mu}{M} = \left( X_\mu ,M\right) 
\label{defV}
\end{equation}
Further quantities $\gamma_\mu$ may analogously be defined as
derivatives of canonical positions $x_\mu$ 
\begin{eqnarray}
&&\gamma _\mu  = \left( x_\mu ,M\right) 
= V_\mu - 2 \gamma \frac{S_\mu} {\hbar}
\qquad S_\mu = \frac{W_\mu} {M} 
\label{defCliff}
\end{eqnarray}
Notice that $S_\mu$ and $\gamma$ anticommute.

Two velocities have been defined which have quite different
properties. The velocities (\ref{defV}) defined from hermitian positions
have the standard form of mechanical velocities with the mass being
however treated as a quantum observable. Their different components
commute with each other.
The velocities $\gamma _\mu$ defined from canonical positions 
involve the mechanical velocities as well as spin terms. 
Their components do not commute but have commutators directly related
to the self-dual spin tensor
\begin{eqnarray}
&&s_{\mu \nu }=-\frac{\hbar ^2 }{4}\left( \gamma _\mu ,\gamma _\nu \right) 
\end{eqnarray}
These velocities also allow to write the mass $M$ as a linear expression
of momenta as in standard Dirac theory
\begin{equation}
M = P^\mu \gamma _\mu = \gamma _\mu P^\mu  \label{Dirac}
\end{equation}
The two foregoing relations suggest that the
velocities $\gamma_\mu $ are the extensions in our quantum 
algebraic framework of the Clifford matrices of Dirac theory. 

This statement can effectively be put on firm grounds.
To this aim, we evaluate the component of the spin vector
$S_\mu$ measured along an arbitrary unit vector $n^\mu$ 
transverse to momentum 
\begin{equation}
S_\mu  n^\mu = - \frac{\hbar} 2 \gamma \gamma_\mu n^\mu 
\end{equation}
For a spin $\frac 12$, this component can only take the two values 
$\pm \frac {\hbar} 2 $. This spectral condition may also be written 
as a characteristic polynomial 
\begin{equation}
S_\mu \cdot S_\nu = - \frac{\hbar ^2 }{4}\left( \eta _{\mu \nu }-\frac{P_\mu 
P_\nu }{M^2 }\right)   \label{spinonehalf}
\end{equation}
One deduces that the velocities (\ref{defCliff}) 
generate a Clifford algebra with $4$ generators  
\begin{eqnarray}
&&\gamma _\mu \cdot \gamma _\nu =\eta _{\mu \nu }
\label{Clifford}
\end{eqnarray}
One then derives that they commute with momenta and canonical positions 
\begin{eqnarray}
&&\left( P_\mu ,\gamma _\nu \right) =
\left( x_\mu ,\gamma _\nu \right) = 0
\end{eqnarray}
These results show how the velocities $\gamma_\mu $ may be considered 
as quantum algebraic extensions of Clifford matrices. 
Notice that the mass $M$ is now a quantum operator 
which anticommutes with $\gamma$ and has a non vanishing commutator 
with $D$, in consistency with the appropriate dimension of a mass. 
Hence this operator is distinct from the classical mass constant 
$m$ of Dirac theory and, accordingly, the Clifford generators (\ref{defCliff}) 
cannot be confused with Clifford matrices of Dirac theory. 
Clifford matrices are fundamental entities in standard Dirac theory.
Here, the expression (\ref{defCliff}) of Clifford generators 
has been derived from a few basic assumptions associated with
symmetry principles and fermionic character of electrons.

A few remarks are worth of consideration at this point.
First the mass $M$ has different signs for electrons and positrons 
so that its modulus $\left| M\right|$ rather than itself has to be 
interpreted as the quantum counterpart of the classical constant $m$.
Then spin terms cannot be considered as small corrections
in (\ref{defCliff}) since Clifford velocities have a magnitude always equal 
to the velocity of light, due to (\ref{Clifford}), whereas mechanical 
velocities are usually much smaller than the velocity of light. 
Finally $\gamma $ may be written as the product of Clifford
generators and it anticommutes with each of them 
\begin{equation}
\gamma =i\gamma _0\gamma _1 \gamma _2 \gamma _3
\qquad \gamma \cdot \gamma _\mu = 0
\end{equation}
  
The description of electrons presented here has been written down in a purely
quantum algebraic manner. In this respect, it presents interesting 
analogies with the description of leptons in Connes' non commutative 
geometry \cite{Connes94} where the ``Dirac operator'' $P^\mu \gamma _\mu$
and the involution $\gamma$ also play primary roles.
Notice however that the present description is built on 
fundamental symmetry principles embodied in the conformal algebra.
The expressions (\ref{defCliff}) of Clifford 
generators, the Dirac relation (\ref{Dirac}) and the Clifford properties 
(\ref{Clifford}) have not been assumed but rather derived from symmetries.
Moreover, the effects of acceleration will be derived in the next section 
from the same principles and not postulated as a further separate assumption.

\section{Accelerated frames}

We now aim to describe quantum electrons not only in inertial frames but 
also in uniformly accelerated frames.

As a first step, we complete the conformal algebra by
considering the generators of transformations to accelerated 
frames and adding the following commutation relations to the already
known ones (\ref{PJ},\ref{PJD}) 
\begin{eqnarray}
&&\left( P_\mu ,C_\nu \right) =-2\eta _{\mu \nu }D-2J_{\mu \nu } 
\nonumber \\
&&\left( J_{\mu \nu },C_{\rho }\right) =\eta _{\nu \rho }C_\mu -\eta _{\mu
\rho }C_\nu   \nonumber \\
&&\left( D,C_\mu \right) =-C_\mu \qquad \left( C_\mu ,C_\nu \right)
=0  \label{PJDC}
\end{eqnarray}
The quantities $C_\mu $ generate relativistic shifts under infinitesimal 
transformations to uniformly accelerated frames. In particular, the first
line of (\ref{PJDC}) gives the redshift laws for energy-momentum operators.
As already discussed, the foregoing relations also represent the quantum
commutation rules between $C_\mu $ and other observables. 
Then, finite frame transformations are described by conjugations in the 
group built on the conformal algebra. Precisely, the shift of an
observable from $A$ in a frame to $\overline{A}$ in another one is read,
for transformations to accelerated frames, as
\begin{equation}
\overline{A}=\exp \left( -\frac{\alpha ^{\rho }C_{\rho }}{i\hbar }\right)
A\exp \left( \frac{\alpha ^{\rho }C_{\rho }}{i\hbar }\right)   \label{deftra}
\end{equation}
The parameters $\alpha ^{\rho }$ are classical accelerations along the $4$
space-time directions. 

Clearly, this conjugation preserves the structure
of quantum algebraic relations known in inertial frames. For example,
position and momentum observables are transformed under (\ref{deftra})
but the canonical commutators between them are preserved
since $\eta _{\mu \nu }$ is a classical number invariant under conjugations. 
Canonical commutators have thus the same form in accelerated and
inertial frames and can be written in terms of the same Minkowski metric.
This result had to be expected in a quantum algebraic approach but it
clearly stands in contradistinction with covariant techniques of classical
relativity. However the metric properties of classical relativity will be 
recovered or, more properly, generalized to a quantum algebraic framework
in the forthcoming developments based on the evaluation of
the redshifts of observables under the group conjugation (\ref{deftra}).

Explicit expressions of the redshifts obviously depend on the relations 
existing between generators $C_\mu $ on one hand, Poincar\'{e} and 
dilatation generators on the other hand. 
A general study of such relations, relying on general properties
of the conformal algebra, is already available \cite{EPJD98}. 
We obtain now more specific results by assuming that
electrons are fundamental particles with a spin number $s=\frac1 2 $ 
preserved under frame transformations. Preservation of the spin
number entails that generators $C_\mu $ have closed expressions 
in terms of Poincar\'{e} and dilatation generators \cite{EPJD98}. 
These expressions take a simple form when written with canonical 
variables 
\begin{equation}
C_\mu =2D\cdot x_\mu -P_\mu \cdot x^2 +2x^{\rho }\cdot s_{\rho \mu }
\label{defC}
\end{equation}

We may now derive redshifts (\ref{deftra}) through straightforward 
algebraic computations, using the expression (\ref{defC}) of $C_\mu $ 
and the simple commutation relations of canonical variables.
As a first important output, the mass $M$ is found to vary as a
position-dependent conformal factor 
\begin{eqnarray}
\overline{M}=M\cdot \frac1 {\lambda } &\qquad &\frac1 {\lambda }=1-2\alpha^\mu
x_\mu +\alpha ^2 x^2   \nonumber \\
M=\overline{M}\cdot \lambda  &\qquad &\lambda =1+2\alpha^\mu \overline{x}_\mu%
+\alpha ^2 \overline{x}^2   \label{traM}
\end{eqnarray}
The conformal factor $\lambda$ also appears in the transformation of canonical
positions which has the form expected from classical relativity 
\begin{equation}
\overline{x}^\mu =\lambda \left( x^\mu -x^2 \alpha ^\mu \right)
\qquad x^\mu =\frac1 {\lambda }\left( \overline{x}^\mu +\overline{x}%
^2 \alpha ^\mu \right)   \label{traKsi}
\end{equation}
Its interpretation as a conformal factor is confirmed by the relation
between the metric tensors evaluated in the two frames 
\begin{eqnarray}
&&g_{\mu \nu }=\partial _\mu \overline{x}^{\rho }\eta _{\rho \sigma
}\partial _\nu \overline{x}^{\sigma }=\lambda ^2 \eta _{\mu \nu }\qquad
\partial _\mu \equiv \frac{\partial }{\partial x^\mu }  \nonumber \\
&&\overline{g}_{\mu \nu }=\eta _{\mu \nu }  \label{traG}
\end{eqnarray}
To fix ideas, we have supposed the coordinate frame $\overline{x}^\mu $ to
be inertial and the coordinate frame $x^\mu $ to be accelerated. 
We however emphasize that the description of frame transformations by equations 
(\ref{traM}-\ref{traG}) is completely reciprocal in accordance 
with the group structure embodied in (\ref{deftra}). 

The mass (\ref{traM}) has the same expression as in conformally invariant
generalizations of Dirac theory \cite{SchoutenH,Dirac36,Fierz39,Pauli40}.
But such generalizations were based on an {\it adhoc} prescription where
mass is a classical parameter varying according to the classical metric
factor. This is why these generalizations were never able to reach the
standards of full quantum consistency. The same remark holds for
generalizations of Dirac theory to Riemannian space 
\cite{Schouten31,Schrodinger32,Weinberg72}. The results obtained here have a
quite different status since they have been derived in a consistent quantum
framework. In fact, the metric factor has been derived from the variation
of mass and it has been obtained as a function of quantum
canonical positions. Therefore this metric factor is itself a 
quantum observable. 

Using the shift laws obtained for $M$ and $x$, we deduce the transformation
of other quantities, in particular of the tetrad of Clifford generators 
\begin{eqnarray}
\overline{\gamma }_\mu &=&\lambda e_\mu ^{\ \nu }\gamma _\nu 
\nonumber \\
e_\mu ^{\ \nu } \left( x\right) &=& \overline{\partial }_\mu x^\nu 
=\frac1 {\lambda ^2}\left( \partial ^\nu  \overline{x}_\mu  \right) 
\label{traE}
\end{eqnarray}
This transport law, written in terms of a {\it vierbein} matrix 
$e_\mu ^{\ \nu }$, ensures that Clifford relations (\ref{Clifford}) 
are preserved under frame transformations as well as the orientation 
$\gamma$ of the tetrad. 

The redshift of energy-momentum is then seen to involve a
spin dependent part 
\begin{equation}
\overline{P}_\mu =e_\mu ^{\ \nu }\cdot P_\nu +\frac1 2 \left(
\partial ^{\rho }e_\mu ^{\ \nu }\right) s_{\nu \rho }  \label{traP}
\end{equation}
This means that translations are transformed as a covariant vector provided
that a spin term is added which has the form of a connection
\cite{deWitt63,Weinberg72}. However, the connection appearing in (\ref
{traP}) is a quantum operator written as a function of canonical position
and spin variables. Moreover, its expression is not a further assumption but
an output of conformal algebra.

Expressions (\ref{traM}-\ref{traP}) have a simple form when
written in terms of canonical variables but, as already noticed,
they involve non hermitian operators. Alternatively,
the shifts may be written in terms of hermitian observables $X_\mu $ and 
$S_{\mu \nu }$, and then involve momentum dependent corrections 
\begin{eqnarray}
&&\overline{M}=M\cdot \left(
1-2\alpha^\mu X_\mu +\alpha^2 \left( X^2 + 
\frac{3 \hbar^2}{4 P^2} \right) \right) \nonumber\\ 
&&\overline{P}_\mu =E_\mu ^{\ \nu }\cdot P_\nu +\frac1 2 \partial
^{\rho }E_\mu ^{\ \nu }\cdot S_{\nu \rho }+\frac{3 \hbar^2}{32} \partial _{\nu
}\partial ^{\rho }E_\mu ^{\ \nu }\cdot \frac{P_{\rho }}{P^2 }  \nonumber\\
&&E_\mu ^{\ \nu }=e_\mu ^{\ \nu }\left( X\right) 
\label{traP_XS}
\end{eqnarray}
The function $e_\mu ^{\ \nu }$ being defined from the tetrad 
transformation (\ref{traE}) has been used to build $E_\mu ^{\ \nu }$
through a substitution of hermitian positions $X$ to canonical ones $x$. 
Since this function is a second-order polynomial form, an ordering
has to be chosen when writing it. However, this ordering does not 
matter in the expression of $\overline{P}_\mu $ due to the form 
(\ref{XX}) of the commutators between hermitian positions.

\section{Discussion}

A new quantum algebraic description of electrons has been presented
in this paper. This description fulfills the relativistic and
quantum requirements discussed in the introduction in a completely
consistent theoretical framework.
We have not specified the quantum field theory except for the basic
properties that it is a conformally invariant description of spin
$\frac 12$ fermions. The definition of localization observables,
the expression of Clifford generators, the evaluation of quantum 
commutators and relativistic shifts have all been derived within 
a single calculus built on conformal algebra. 

Frame transformations have been described as group conjugations. This 
ensures that the quantum algebraic relations defined in inertial frames 
may be exported to uniformly accelerated frames. Although it is quite
different from covariance techniques of classical relativity, this
description has allowed to recover in a quantum framework a lot
of geometric laws known from classical relativity.
Mass has been found to experience a redshift which fits
the expectation deduced from Einstein equivalence principle. 
This mass redshift may be considered as defining a quantum conformal
factor which has also been shown to enter the expression of a quantum
metric tensor.
Expressions obtained for other observables not only 
translate the geometric laws of classical relativity into a quantum 
theoretical framework but they also change this laws through the 
addition of spin terms. 

These results should open the way to a renewed treatment of 
effects of acceleration or gravitation on quantum systems 
which could in particular show useful for analysing high precision
tests of inertial or gravitational effect on atomic clocks 
\cite{IEEE91}.

\end{multicols}

\end{document}